\documentclass[a4paper,fleqn,usenatbib]{mnras}
\usepackage{newtxtext, newtxmath}
\usepackage[T1]{fontenc}
\usepackage{ae,aecompl}

\usepackage{graphicx}	
\usepackage{amsmath}	
\usepackage{amssymb}	
\graphicspath{{./}{figures/}}

\graphicspath{{./}{figures/}}

\title[]{A Three Dimensional View of Gomez's Hamburger}

\author[]
{\parbox{\textwidth}{Richard Teague$^{1}$, Marija R. Jankovic$^{2}$, Thomas J. Haworth$^{3}$, Chunhua Qi$^{1}$, John D. Ilee$^{4}$
}\vspace{0.4cm}\\
\parbox{\textwidth}{
$^{1}$ Center for Astrophysics | Harvard \& Smithsonian, 60 Garden Street, Cambridge, MA 02138, USA\\
$^{2}$ Astrophysics Group, Imperial College London, Blackett Laboratory, Prince Consort Road, London SW7 2AZ, UK\\
$^{3}$ Astronomy Unit, School of Physics and Astronomy, Queen Mary University of London, London E1 4NS, UK\\
$^{4}$ School of Physics and Astronomy, University of Leeds, Leeds LS2 9JT, UK
}}

\begin{document}
\date{Accepted XXX. Received YYY; in original form ZZZ}

\pubyear{2020}


\label{firstpage}
\pagerange{\pageref{firstpage}--\pageref{lastpage}}
\maketitle

\begin{abstract}
Unraveling the 3D physical structure, the temperature and density distribution, of protoplanetary discs is an essential step if we are to confront simulations of embedded planets or dynamical instabilities. In this paper we focus on Submillimeter Array observations of the edge-on source, Gomez's Hamburger, believed to host an over-density hypothesised to be a product of gravitational instability in the disc, GoHam~b. We demonstrate that, by leveraging the well characterised rotation of a Keplerian disc to deproject observations of molecular lines in position-position-velocity space into disc-centric coordinates, we are able to map out the emission distribution in the $(r,\,z)$ plane and ($x,\, |y|,\, z)$ space. We show that $^{12}$CO traces an elevated layer of $z\,/\,r \sim 0.3$, while $^{13}$CO traces deeper in the disc at $z\,/\,r \lesssim 0.2$. We identify an azimuthal asymmetry in the deprojected $^{13}$CO emission coincident with GoHam~b at a polar angle of $\approx 30\degr$. At the spatial resolution of $\sim 1.5\arcsec$, GoHam b is spatially unresolved, with an upper limit to its radius of $<190$~au.
\end{abstract}


\section{Introduction}

\begin{figure*}
    \centering
    \includegraphics[width=\textwidth]{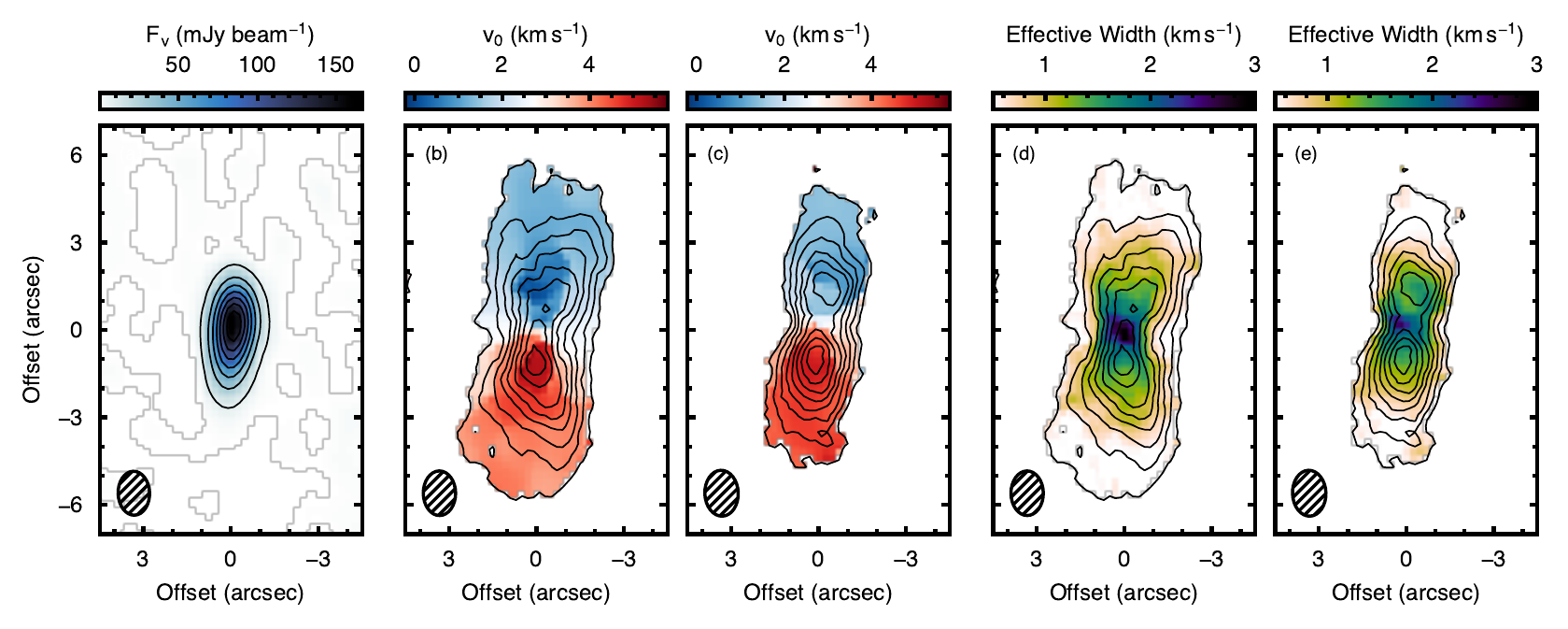}
    \caption{Summary of the observations. Panel (a) shows the 1.3~mm continuum emission. The black contours show steps of $20\sigma$ starting at $10\sigma$, where $\sigma = 1.13~{\rm mJy\,beam^{-1}}$. Central panels (b) and (c) show the rotation maps for the $^{12}$CO and $^{13}$CO emission, respectively. The black contours show the integrated intensities for the two lines in steps of 10\% of their peak values, $4.90~{\rm Jy\,beam^{-1}\,km\,s^{-1}}$ and $3.51~{\rm Jy\,beam^{-1}\,km\,s^{-1}}$. The right most panels, (d) and (e) show the effective width of the line, with the integrated intensity contours overlaid. The synthesized beams are shown in the bottom left of each panel.}
    \label{fig:GoHam_observations}
\end{figure*}

High angular resolution observations of the dust in protoplanetary discs, both at mm and NIR wavelengths, have shown a stunning variety of features such as concentric rings and spirals \citep{Andrews_ea_2018, Avenhaus_ea_2018}. These structures hint at highly dynamic environments where the dust distributions are sculpted by changes in the gas pressure distribution. The precise cause for the perturbations in the gas is hard to constrain, with multiple scenarios possible including embedded planets \citep[e.g.][]{2015MNRAS.453L..73D, 2018A&A...617A..44K, 2018A&A...610A..24F, Zhang_ea_2018}, (magneto-)hydrodynamical instabilities \citep{Flock_ea_2015} or gravitational instabilities \citep{2015MNRAS.451..974D, 2015ApJ...812L..32D, Hall_ea_2016, 2017ApJ...839L..24M}. Differentiating between these scenarios requires an intimate knowledge of the underlying gas structure and, in particular, how that structure changes from the midplane, as traced by the mm continuum emission, to the disc atmosphere, populated by the small sub~$\micron{}$ grains which efficiently scatter stellar NIR radiation.

This is routinely attempted by using observations of different molecular species believed to trace distinct vertical regions in the disc. This is due to a combination of both optical depth effects and changes in physical conditions with height in the disc which make certain regions more conducive to the formation of particular species. However, it is only with high spatial resolution data that we are begining to be able to directly measure the height at which molecular emission arises \citep{Rosenfeld_ea_2013, deGregorio-Monsalvo_ea_2013, Pinte_ea_2018}, verifying predictions from chemical models.

A more direct approach is the observation of high inclination discs where the emission distribution can be mapped directly. Unlike continuum emission which suffers from extremely high optical depths due to the long path lengths for edge on discs \citep{Guilloteau_ea_2016, Louvet_ea_2018}, the rotation of the disc limits the optical depth of molecular emission in a given spectral channel. This allowed \citet{Dutrey_ea_2017} to map the $^{12}$CO $J = 2-1$ and CS $J = 5-4$ emission distribution in the $(r_{\rm disc},\, z_{\rm disc})$ plane, calling this a tomographically reconstructed distribution, for the edge-on disc colloquially known as the Flying Saucer (2MASS J16281370-2431391).

In addition to allowing access to the $(r_{\rm disc},\, z_{\rm disc})$ plane, \citet[]{Dent_ea_2014}, but see also \citet{Matra_ea_2017} and \citet{Cataldi_ea_2018}, demonstrated how similar techniques can be used to deproject a cut across the disc major axis into the $(x_{\rm disc},\, |y_{\rm disc}|)$ plane. The absolute value of $y_{\rm disc}$ arises because it is impossible to distinguish between the near and far side of the disc ($\pm y_{\rm disc}$) from their projected line of sight velocities alone. Using this technique, the authors were able to extract the azimuthal emission distribution along the line of sight revealing a clump of CO emission. Application of this technique to a vertically extended source enables the extraction of a full 3D emission distribution.

In this paper we apply these techniques to Submillimeter Array (SMA) observations of Gomez's Hamburger, an edge-on circumstellar disc. In section~\ref{sec:observations} we describe the observations and data reduction. In Section~\ref{sec:deprojection} we provide an overview of the deprojection techniques used and their application to Gomez's Hamburger. A discussion of these results and a summary conclude the paper in Sections~\ref{sec:discussion} and \ref{sec:summary}, respectively.

\section{Summary of Observations}
\label{sec:observations}

\begin{figure*}
    \centering
    \includegraphics[width=\textwidth]{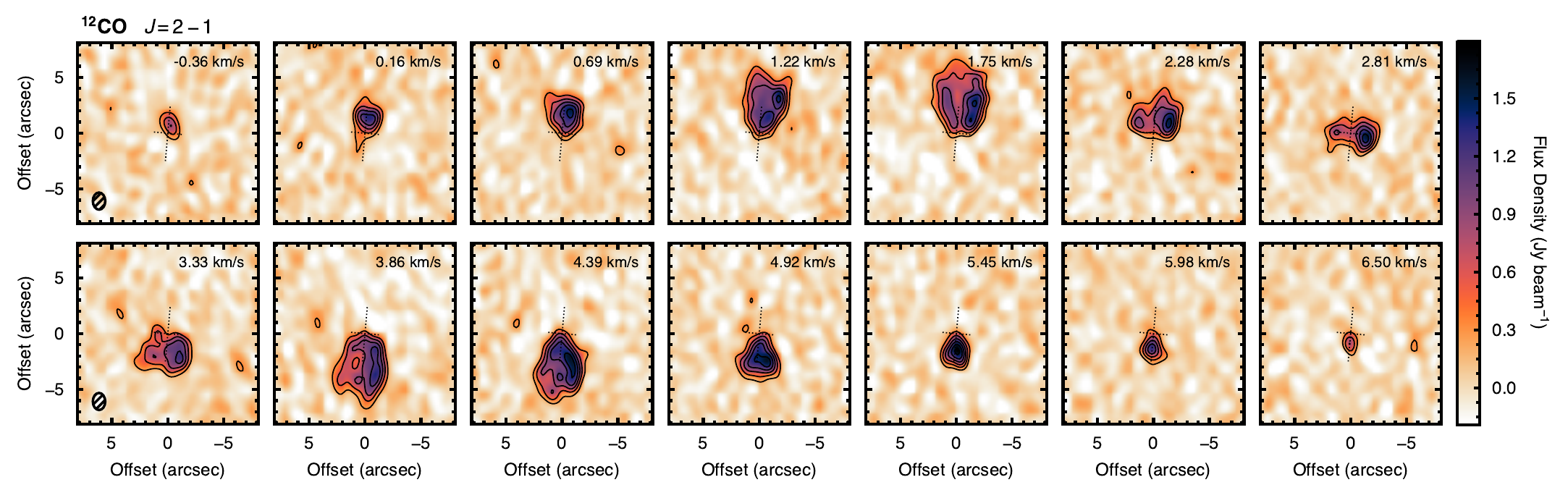}\vspace{0.5cm}
    \includegraphics[width=\textwidth]{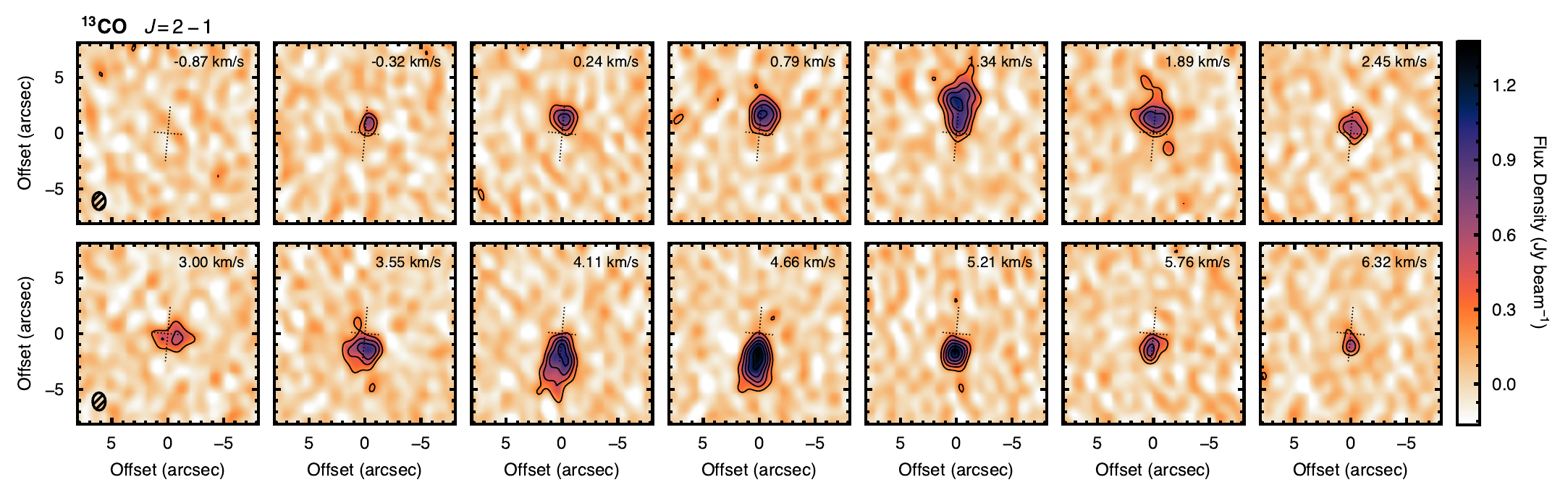}
    \caption{\emph{Top}: Channel maps of the $^{12}$CO emission, downsampled to 528~${\rm m\,s^{-1}}$ channel spacing for presentation. Solid lines show contours starting at $3\sigma$ and increasing in steps of $3\sigma$, where $\sigma = 87~{\rm mJy\,beam^{-1}}$. The synthesized beams are shown in the left rows. The dotted lines show the orientation of the major and minor axes of the disc. The central velocity of the channel is shown in the top right of each panel. \emph{Bottom}: As above, but for $^{13}$CO emission. The solid lines are contours of $3\sigma$ where $\sigma = 75~{\rm mJy\,beam^{-1}}$. Note the substantial increase in brightness in the southern side of the disc.}
    \label{fig:GoHam_channels}
\end{figure*}

At an inclination of $i \approx$\,$86\degr$ and a distance of $250 \pm 50$~pc, Gomez's Hamburger (GoHam, IRAS~18059-3211) offers a rare opportunity to study the chemical and physical structure of an edge-on disc. Although originally classified as an evolved A0 star surrounded by a planetary nebula, follow-up observations using the Submillimeter Array (SMA) showed CO emission in the distinct pattern of Keplerian rotation about GoHam. These and subsequent observations firmly establish GoHam as a $2.5 \pm 0.5$ M$_{\odot}$ A-type star at a distance of $250 \pm 50$~pc surrounded by a massive, $M_{\rm disc} \sim 0.2~M_{\rm sun}$, circumstellar disc \citep{Bujarrabal_ea_2008, Bujarrabal_ea_2009, Wood_ea_2008, deBeck_ea_2010}. This identification is further justified with the exquisite observations from the NICMOS instrument on the Hubble Space Telescope (HST) which show the distinct flared geometry associated with protoplanetary discs \citep{Bujarrabal_ea_2009}.

\subsection{Data Reduction}

The data were obtained from the SMA archive\footnote{\url{https://www.cfa.harvard.edu/cgi-bin/sma/smaarch.pl}} and calibrated using the \texttt{MIR} software\footnote{\url{https://www.cfa.harvard.edu/~cqi/mircook.html}}. The interested reader is referred to the original papers, \citet{Bujarrabal_ea_2008, Bujarrabal_ea_2009}, for a thorough overview of the calibration process. After calibration, the data were exported to \texttt{CASA v5.6.0} where two rounds of self-calibration were performed on the continuum, with phase-solutions applied to the spectral line windows. The phase center was adjusted so that the center of the continuum was in the image center.

After experimenting with various imaging properties, both the $^{12}$CO and $^{13}$CO transitions were imaged at their native channel spacing of $203$~kHz ($264~{\rm m\,s^{-1}}$) with a Briggs weighting scheme and a robust parameter of 0.5. This resulted in synthesized beams of $1.53\arcsec \times 1.11\arcsec$ at 0.4$\degr$ for $^{12}$CO and $1.57\arcsec \times 1.15\arcsec$ at $2.0\degr$ for $^{13}$CO. The measured RMS in a line free channel was found to be $132~{\rm mJy~beam^{-1}}$ and $120~{\rm mJy~beam^{-1}}$ for the $^{12}$CO and $^{13}$CO. Channel maps were created both at the native channel spacing and downsampled by a factor of two to increase the signal to noise ratio. 

Moment maps were also generated for the data using the Python package \texttt{bettermoments} \citep{Teague_Foreman-Mackey_2018}. Integrated intensity maps were created using a threshold of $2\sigma$ for both molecules, while the rotation map used the quadratic method described in \citep{Teague_Foreman-Mackey_2018} without the need for any $\sigma$-clipping. Rather than using the intensity weighted velocity dispersion (second moment) which is typically very noisy and incurs a large uncertainty \citep{Teague_2019}, we use the `effective line width' implemented in \texttt{bettermoments}. This calculates an effective line width using $\Delta V_{\rm eff} = M_0 / \sqrt{\pi} \, F_{\nu}^{\rm max}$, where $M_0$ is the integrated intensity and $F_{\nu}^{\rm max}$ is the line peak. For a Gaussian line profile, this returns the true Doppler width of the line. Both transitions show a peak at the disc center, gradually decreasing in the outer disc. However, at this spatial resolution the line profile is dominated by systematic broadening effects from the imaging.

\subsection{Observational Results}

Using the 2D-Gaussian fitting tool \texttt{IMFIT} in \texttt{CASA} the integrated flux of the 1.3~mm continuum was found to be $293 \pm 4~{\rm mJy}$, consistent with \citet{Bujarrabal_ea_2008}. Integrating over an elliptical region with a major axis of 14\arcsec{}, a minor axis of 7\arcsec{} and a position angle of 175\degr{}, and clipping all values below $2 \sigma$, the $^{12}$CO integrated flux was found to be 37.2~${\rm Jy\,km\,s^{-1}}$. For the $^{13}$CO, integrating over an elliptical mask with a major axis of 12\arcsec{} and a minor axis of 4.2\arcsec{} a position angle of 175\degr{}, again clipping all values below $2 \sigma$, resulted in an integrated flux of 16.5~${\rm Jy\,km\,s^{-1}}$.

A summary of the moment maps alongside the continuum image is shown in Fig.~\ref{fig:GoHam_observations}. The continuum is clearly detected and considerably smaller in extent than the gas component. Assuming a source distance of $250$~pc \citep{Bujarrabal_ea_2008}, the gaseous disc extends 1500~au in radius. For both transitions the southern side of the disc is observed to be considerably brighter than the northern side, in addition to a slight north-south asymmetry in the continuum emission. In addition, the east-west asymmetry in the $^{12}$CO integrated intensity suggests that the eastern side of the disc is tilted towards the observer. 

Figure~\ref{fig:GoHam_channels} shows the channel maps, downsampled in velocity by a factor of two, for the $^{12}$CO emission, top, and the $^{13}$CO emission, bottom. Both lines show the distinct `butterfly' emission morphology characteristic of a rotating disc. The $^{12}$CO emission is more extended, both in the radial and vertical directions, as would be expected given its larger abundance. The $^{12}$CO emission also splits into two lobes, most clearly seen in the channels at $1.62~{\rm km\,s^{-1}}$ and $3.73~{\rm km\,s^{-1}}$, due to the elevated emission surface, while the $^{13}$CO appears more centrally peaked.

To find the systemic velocity of the disc, we fit the rotation maps, maps of the line center, $v_0$, shown in Fig.~\ref{fig:GoHam_observations}, using the Python package \texttt{eddy} \citep{eddy}. At these large inclinations, vertically extended emission, as expected for $^{12}$CO and to a lesser extent, $^{13}$CO, will result in rotation maps which are extended along the minor axis \citep[see Fig.~3a from][]{Dutrey_ea_2017}, resulting in a $v_0$ distribution which deviates significantly from an inclined 2D disc model. Despite this, the rotation profile will be symmetric about the systemic velocity such that the inferred $v_{\rm LSR}$ from a fit of an inclined 2D disc will provide a good estimate of the true systemic velocity. We fix the source distance to 250~pc, and allow the source center, inclination, position angle, stellar mass and systemic velocity to vary. Using 64 walkers which take 10,000 burn-in steps and an addition 5,000 steps to estimate the posterior distributions, we find Gaussian-like posteriors for $v_{\rm LSR}$ for both transitions: $v_{\rm LSR}(^{12}{\rm CO}) = 2793 \pm 34~{\rm m\,s^{-1}}$ and $v_{\rm LSR}(^{13}{\rm CO}) = 2778 \pm 38~{\rm m\,s^{-1}}$. These uncertainties represent the statistical uncertainties which do not consider the applicability of the model and so the true uncertainties are likely larger.

\section{Deprojection to Disc-Centric Coordinates}
\label{sec:deprojection}

\begin{figure}
    \centering
    \includegraphics[width=\columnwidth]{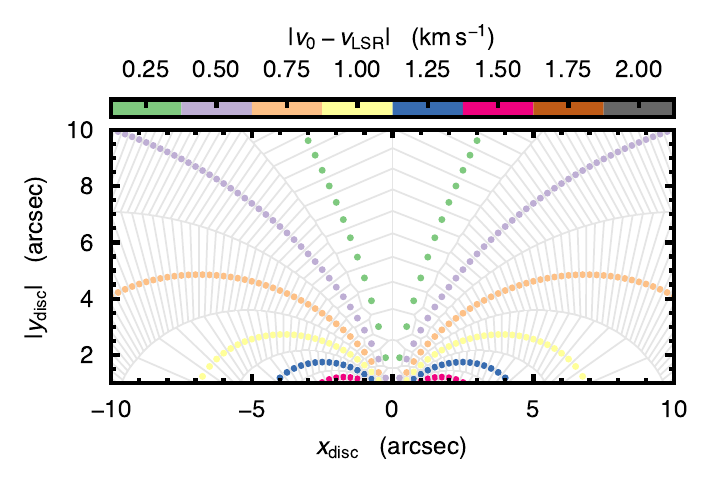}
    \caption{The deprojection of pixels assuming a 0.25\arcsec{} pixel size and a $250~{\rm m\,s^{-1}}$ velocity spacing from Eqn.~\ref{eq:y}. The velocity resolution sets the number of `spokes' in the deprojection, while the pixel scaling sets the sampling along each spoke. The central channels close to the systemic velocities, $v_0 \sim v_{\rm LSR}$, populate regions where $x_{\rm disc}$ is small. Thus, even with high velocity resolution, it is hard to get an accurate deprojection for these regions \citep[see, for example:][]{Dent_ea_2014, Matra_ea_2017, Cataldi_ea_2018}.}
    \label{fig:pixel_deprojection}
\end{figure}

If the velocity structure of the source is known, it is possible to deproject observations of an edge-on disc in position-position-velocity (PPV) space, $(x_{\rm sky},\, y_{\rm sky},\, v_0)$, into 3D disc-centred coordinates, $(x_{\rm disc},\, y_{\rm disc},\, z_{\rm disc})$. Both \citet{Dutrey_ea_2017} and \citet{Matra_ea_2017} discuss similar deprojections, the former into an azimuthally averaged $(r_{\rm disc},\, z_{\rm disc})$ plane, and the latter into the $(x_{\rm disc},\, |y_{\rm disc}|)$ plane for a cut at a constant $z_{\rm disc}$ through the disc. In this section we discuss both deprojections and include a correction due to changes in the rotation velocity as a function of height rather than assuming cylindrical rotation.

At any given voxel (a pixel in PPV space), the projected line of sight velocity, $v_0$, is given by,

\begin{equation}
    \label{eq:v0}
    v_0 = v_{\phi} \, \cos \phi \, \sin i + v_{\rm LSR},
\end{equation}

\noindent where $v_{\phi}$ is the rotation velocity, $\phi$ is the azimuthal angle (not to be confused with the polar angle which is measured in the sky-plane rather than the disc-plane), $i$ is the disc inclination and $v_{\rm LSR}$ is the systemic velocity. For Keplerian rotation we know that,

\begin{equation}
    \label{eq:vkep}
    v_{\phi}(r_{\rm disc},\, z_{\rm disc}) = \sqrt{\frac{GM_{\rm star} r_{\rm disc}^2}{(r_{\rm disc}^2 + z_{\rm disc}^2)^{3/2}}}
\end{equation}

\noindent where $r_{\rm disc}$ and $z_{\rm disc}$ are the cylindrical radius and height in the disc, respectively, dropping the disc subscript for brevity. Substituting this into Eqn.~\ref{eq:v0} and noting that for and edge-on disc, such that $i = 90\degr$, $x_{\rm sky} = r_{\rm disc} \cos \phi$ and $z_{\rm disc} = y_{\rm sky}$, then we find,

\begin{equation}
    v_0 - v_{\rm LSR} =  \sqrt{\frac{GM_{\rm star} x_{\rm sky}^2}{(r_{\rm disc}^2 + y_{\rm sky}^2)^{3/2}}}.
\end{equation}

\noindent As both $x_{\rm sky}$ and $y_{\rm sky}$ are readily measured in the image plane, we can rearrange for $r_{\rm disc}$ giving,

\begin{equation}
    \label{eq:R}
    r_{\rm disc} = \sqrt{\left( \frac{GM_{\rm star} x_{\rm sky}^2}{(v_0 - v_{\rm LSR})^2} \right)^{2/3} - y_{\rm sky}^2}.
\end{equation}

\noindent If cylindrical rotation is assumed, i.e. that there is no $z$ dependence in $v_{\phi}$ in Eqn.~\ref{eq:vkep}, the $y_{\rm sky}^2$ correction term vanishes, recovering the result from \citet{Dutrey_ea_2017}.

Noting that $r_{\rm disc} = \sqrt{x_{\rm disc}^2 + y_{\rm disc}^2}$, where $y_{\rm disc}$ is the line-of-sight axis, we can additionally infer something about the line-of-sight distance of the emission,

\begin{equation}
    \label{eq:y}
    |y_{\rm disc}| = \sqrt{\left( \frac{GM_{\rm star} x_{\rm sky}^2}{(v_0 - v_{\rm LSR})^2} \right)^{2/3} - y_{\rm sky}^2 - x_{\rm sky}^2}.
\end{equation}

\noindent as used in \citet{Matra_ea_2017}. However, as there is a degeneracy in the side of the disc the emission arises, $\pm y$, this recovers an average of both sides of the disc. Again, if the cylindrical rotation is assumed, the $y_{\rm sky}^2$ correction term vanishes in Eqn.~\ref{eq:y}. Figure~\ref{fig:pixel_deprojection} shows how pixels would be deprojected into the $(x_{\rm disc},\, |y_{\rm disc}|)$ plane. It illustrates that the velocity resolution sets the `azimuthal' sampling, i.e. how many spokes there are, while the pixel size (or spatial resolution) will set sampling along these spokes. As, such, both spatial and spectral resolution are required for an accurate deprojection of the data.

In addition to the transformation of the coordinates, it is essential to include the Jacobian such that integrated flux in the deprojected maps is conserved. Following Appendix~C of \citet{Cataldi_ea_2018} we find that the Jacobian for the transformation from $(x_{\rm sky},\, y_{\rm sky},\, v)$ to $(r_{\rm disc}, z_{\rm disc},\, v)$, where $z_{\rm disc} = y_{\rm sky}$, is given by,

\begin{equation}
    J_{r_{\rm disc},\,z_{\rm disc}} = \frac{3 x_{\rm sky}\,r_{\rm disc}}{r_{\rm disc}^2 + z_{\rm disc}^2},
\end{equation}

\noindent
which is a dimensionless transformation, meaning that the units are the same as in a channel map \citep[e.g.][]{Dutrey_ea_2017}. Similarly, to transform the sky-plane coordinates into $(x_{\rm disc},\, |y_{\rm disc}|)$ coordinates, we find,

\begin{equation}
    J_{x_{\rm disc},\,y_{\rm disc}} = \frac{3}{2} \sqrt{GM_{\rm star}} \, x_{\rm disc} y_{\rm disc} \, \big(x_{\rm disc}^2 + y_{\rm disc}^2 + z_{\rm disc}^2\big)^{-7/4},
\end{equation}

\noindent
similar to Eqn.~10 in \citet{Cataldi_ea_2018}, but including an addition $z_{\rm disc}^2$ term owing to our definition of $v_{\phi}$. This Jacobian has units of Hz, owing to the change from position-position-velocity space to position-position-position space. After correcting for the change in velocity with an addition ${\rm d}\nu / c$ term, where ${\rm d}$ is the source distance and $\nu$ the frequency of the line, we have the final units of ${\rm W~m^{-2}~sr^{-1}}$, i.e. a radiance along the $z$-axis.

We note that these derivations assume that the disc is completely edge-on and in Keplerian rotation. \citet{Dutrey_ea_2017} showed how changes in the inclination can affect the deprojection. The authors found that for only moderate deviations from edge-on, i.e. $i \gtrsim 80\degr$, the tomographically reconstructed distribution (TRD, see also section \ref{sec:TRD}) provided a good representation of the underlying physical structure. One half of the disc, either where $z > 0$ or $z < 0$, would be brighter, with this brighter half corresponding to the side of the disc which is closer to the observer. In addition, the vertical extent of the emitting layer would broaden in the $z$ direction, before eventually splitting into two distinct arms when $i \lesssim 80\degr$ and the near and far sides of the disc are spatially resolved.

\subsection{Tomographically Reconstructed Distribution}
\label{sec:TRD}

As the disc is expected to be highly inclinated, $i \sim 85\degr$ \citep{Bujarrabal_ea_2008, Bujarrabal_ea_2009}, we use the deprojection techniques described in Section~\ref{sec:deprojection} to explore the three dimensional structure of the disc, starting with the TRD as used for the Flying Saucer in \citet{Dutrey_ea_2017}. We take the geometrical properties inferred from forward modelling a full 3D model presented in \citet{Bujarrabal_ea_2008, Bujarrabal_ea_2009} which assumed Keplerian rotation around a $2~M_{\odot}$ central star and a disc inclined at $85\degr$, observed at a position angle of $175\degr$.

Using Eqn.~\ref{eq:R}, each pixel is deprojected into $(r_{\rm disc},\, z_{\rm disc})$ space, before being binned into bins equal in size to the pixel. In each bin, we take the maximum value, equivalent to collapsing an image cube along the spectral axis by taking the maximum value along each pixel, e.g. a moment 8 map in \texttt{CASA}.

\begin{figure*}
    \centering
    \includegraphics[]{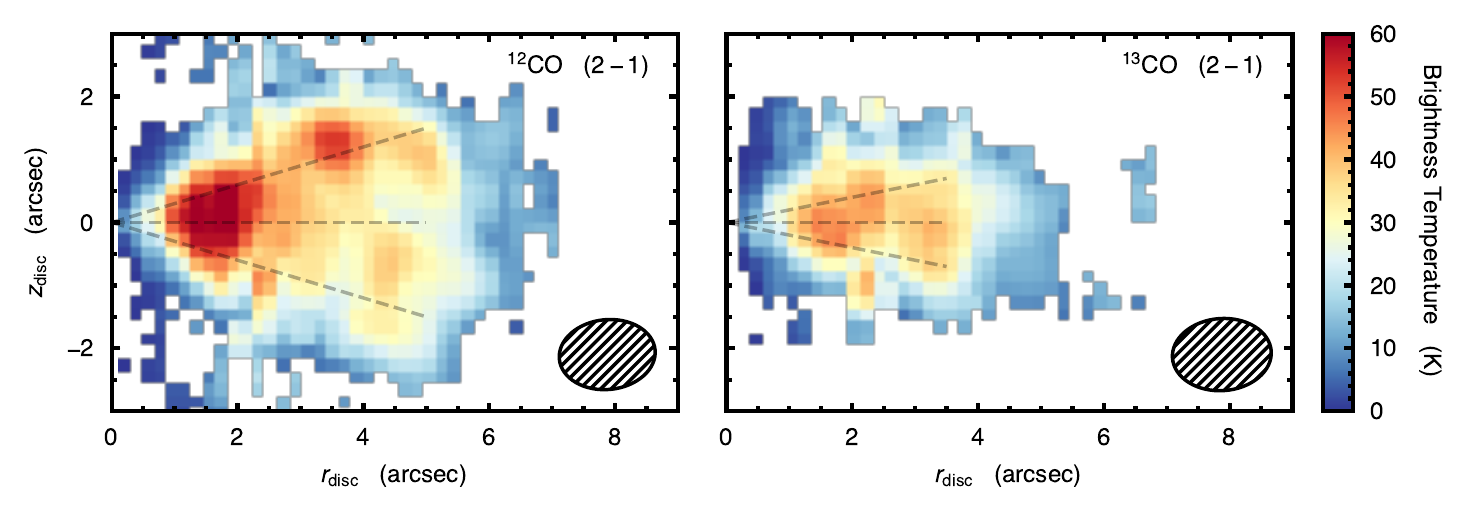}
    \caption{TRD using the method in \citet{Dutrey_ea_2017} for $^{12}$CO, left, and $^{13}$CO, right. The dashed lines show $z \, / \, r = 0.3$ in the left panel and $z \, / \, r = 0.2$ in the right. Note that asymmetry about the $z = 0$ line due to the deviation from a completely edge-on disc as discussed in \citet{Dutrey_ea_2017}.}
    \label{fig:GoHam_TRD}
\end{figure*}

Figure~\ref{fig:GoHam_TRD} shows the TRD for $^{12}$CO, left, and $^{13}$CO, right, taking the peak brightness temperature in each bin. Immediately we see that $^{12}$CO traces an elevated region of $z\,/\,r \sim 0.3$, while $^{13}$CO appears to trace a region closer to the midplane, confined to $z\, / \, r \lesssim 0.2$. The drop off of signal within the inner $1\arcsec$ is due to convolution effects, as described in \citet{Dutrey_ea_2017}. The asymmetry above the midplane is due to the deviation from a directly edge-on disc with a similar effect seen in the Flying Saucer, where the level of difference between the positive and negative values is consistent with the $i \approx 85\degr$ inclination measured for the source. We note a bright point-source at $x_{\rm disc} \sim 3.5\arcsec$ and $z_{\rm disc} \sim 1.5\arcsec$, likely associated with the peak in the $^{12}$CO zeroth moment map in the north-west (see channel 1.22~${\rm km\,s}^{-2}$ in Fig.~\ref{fig:GoHam_cuts_12CO}). Higher resolution data is required to accurately disentangle this feature.

\subsection{Line-of-Sight Deprojection}

In \citet{Bujarrabal_ea_2009} it was argued that there was an enhancement of $^{13}$CO at an offset of $r \approx 1\arcsec$. To explore whether this can be observed with the above techniques, we follow \citet{Matra_ea_2017} and use Eqn.~\ref{eq:y} to deproject cuts along the major axis of the disc into the $(x_{\rm disc},\,|y_{\rm disc}|)$ plane.

\begin{figure*}
    \centering
    \includegraphics[width=\textwidth]{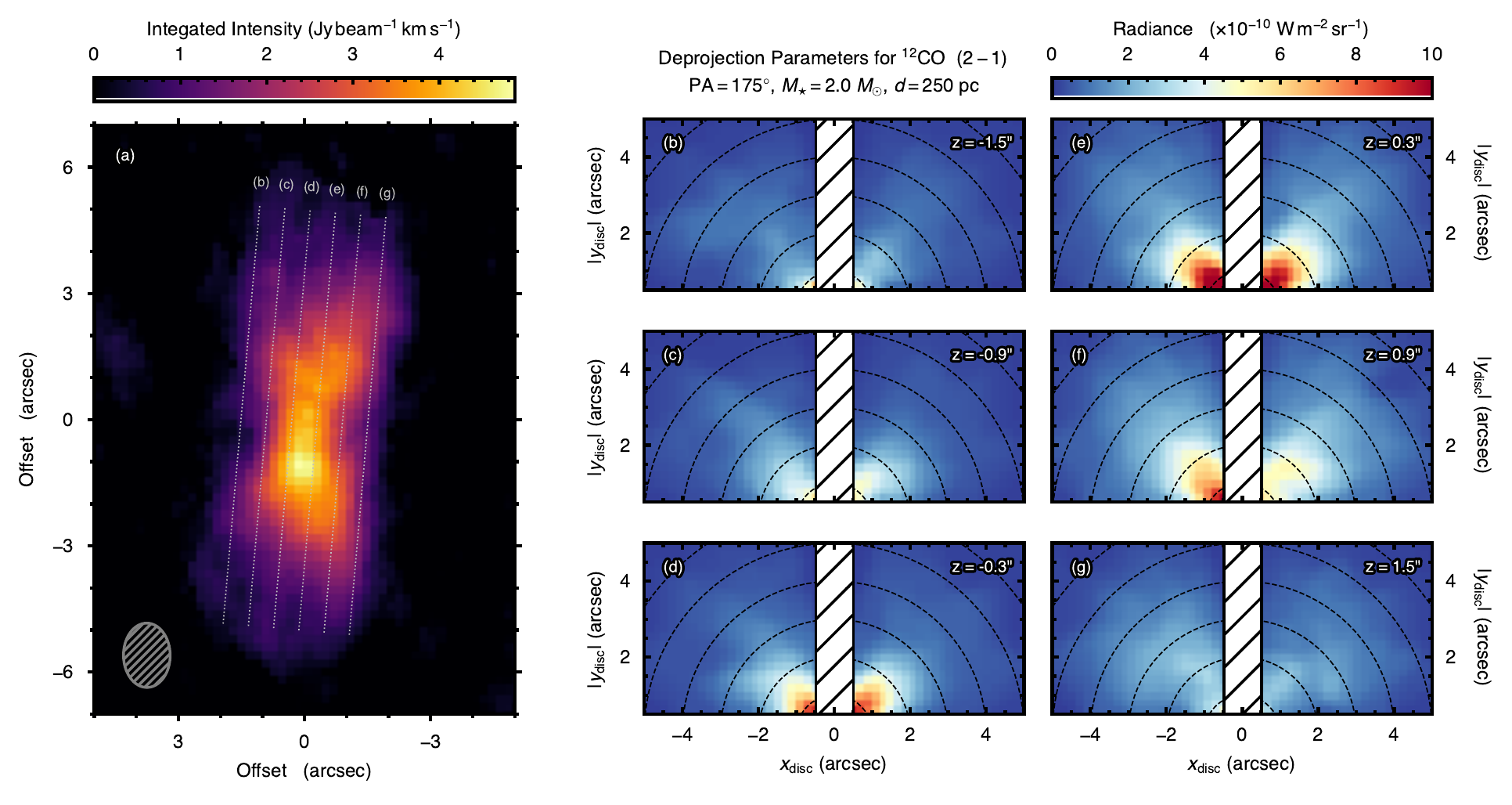}
    \caption{Deprojected $^{12}$CO emission assuming Keplerian rotation. The left panel shows the zeroth moment (integrated intensity) map of $^{12}$CO. The six annotated slices, $(b)$ through $(g)$, show the centre of the cuts which make up the two columns to the right. In the right two columns, $y_{\rm disc}$ represents the line-of-sight axis. For the deprojected data, regions where $|x| < 0.5\arcsec$ and $|y| < 0.5\arcsec$ are masked. The height of each cut relative to the disc midplane is shown in the top right of each panel. Note that negative $x$ values are to the north of the disc center. In panels $(b)$ through $(g)$ the black dashed lines are lines of constant cylindrical radius.}
    \label{fig:GoHam_cuts_12CO}
\end{figure*}

\begin{figure*}
    \centering
    \includegraphics[width=\textwidth]{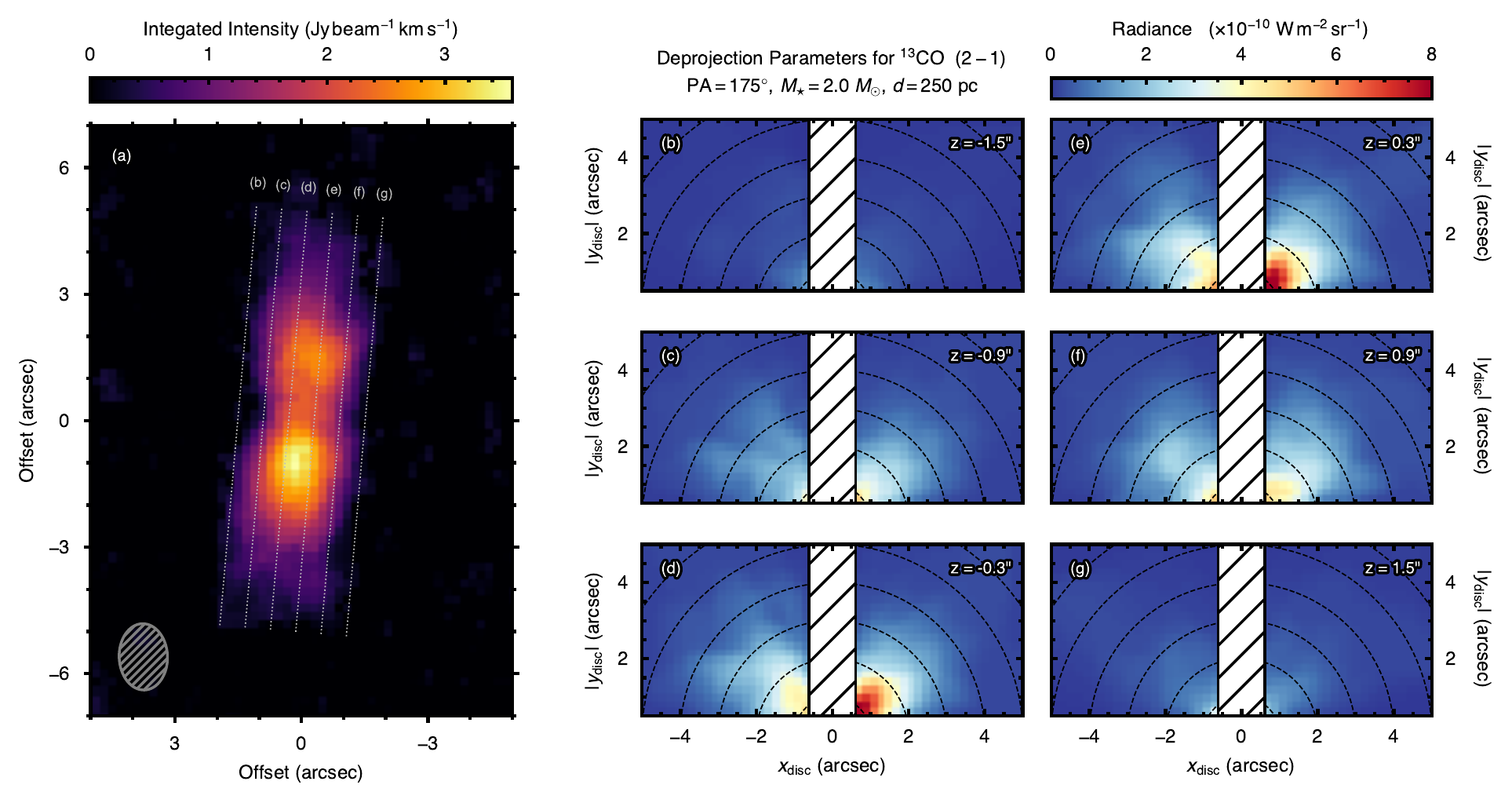}
    \caption{As in Fig.~\ref{fig:GoHam_cuts_12CO}, but for $^{13}$CO. In the integrated intensity map, left, GoHam~b manifests as a bright southern side of the disc. In the right panels, GoHam~b manifests as a significant asymmetry with the positive $x_{\rm disc}$ side of the disc being considerably brighter.}
    \label{fig:GoHam_cuts_13CO}
\end{figure*}

The disc was split into six equally thick slices of $0.8\arcsec$ spanning $\pm 2\arcsec$ about the disc midplane. For each slice, every PPV voxel above a SNR of 2 was deprojected into disc coordinates then linearly interpolated onto a regular grid with the results shown in Fig.~\ref{fig:GoHam_cuts_12CO}. The same procedure was performed for $^{13}$CO, however with narrower slices of $0.6\arcsec$ spanning $\pm 1.5\arcsec$ with the results shown in Fig.~\ref{fig:GoHam_cuts_13CO}.

As with the TRD, the western side of the $^{12}$CO emission, positive $z$ values, panels \emph{(e)}, \emph{(f)} and \emph{(g)}, is considerably brighter than the eastern side, negative $z$ values, due to the slight deviation from a completely edge-on disc \citep{Dutrey_ea_2017}. It is also clear that at large separations from the disc midplane, the inner edge of the $^{12}$CO emission moved outwards, most clearly seen in panels \emph{(b)} and \emph{(g)} of Fig.~\ref{fig:GoHam_cuts_12CO}. Some azimuthal structure is tentatively observed at higher altitudes for $^{12}$CO, namely in panel \emph{(f)}. Given the orientation of the disc, the gas rotates in a clockwise direction. Although the $^{13}$CO data is noisier, some features are still observable. As with the $^{12}$CO, at higher altitudes the emission peaks at $r \sim 3\arcsec$, while becoming more centrally peaks at lower $z$ value. 

Both $^{12}$CO and $^{13}$CO show an enhancement in emission close to the disc midplane, at $(x_{\rm disc}, \, |y_{\rm disc}|,\, z_{\rm disc}) \approx (2\arcsec,\, 1\arcsec,\, 0\arcsec)$, marked in Figures~\ref{fig:GoHam_cuts_12CO} and \ref{fig:GoHam_cuts_13CO} by the black dashed circle. \citet{Bujarrabal_ea_2009} previously reported an enhancement in $^{13}$CO emission at an offset position of $(\delta x_{\rm sky},\, \delta y_{\rm sky}) \approx (1.5\arcsec,\, -2.5\arcsec)$, with later observations of 8.6~\micron{} and 11.2~\micron{} PAH emission  revealing a similar apparent over-density \citep{Berne_ea_2015}. We note that in principle it is possible to subtract an azimuthally averaged model from each of these projected maps. However, we found that given the high azimuthal variability owing to the noise in the data and strong systematic feature due to the transformation from the limited spatial resolution of the data, these did not yield residual maps in which structure was readily distinguished, with higher spatial and spectral resolution data necessary for such an approach.

\section{Discussion}
\label{sec:discussion}

In the previous section we have shown that assuming that an edge-on disc is in Keplerian rotation allows one to deproject pixels in position-position-velocity space into disc-centric position-position-position space. In this section we discuss the implication of these deprojections.

\subsection{GoHam~b}

Previous studies of GoHam have detected a significant enhancement of emission in the southern half of the disc, dubbed GoHam~b, seen in $^{13}$CO emission and 8.6~\micron{} and 11.2~\micron{} PAH emission \citep{Bujarrabal_ea_2009, Berne_ea_2015}. They find that this excess emission could be explained with a gaseous over-density containing a mass of 0.8 to $11.4~M_{\rm Jup}$, spread uniformly over a spherical region with a radius of $\sim 0.6\arcsec$ ($\sim 150$~au). Furthermore, based on models of the disc structure, it is estimated that the disc of GoHam is marginally gravitationally unstable, with Toomre parameter $Q \lesssim 2$ \citep{Berne_ea_2015}. In circumstellar discs gravitational instabilities can lead to growth of local, gravitationally bound over-densities \citep[i.e., to disc fragmentation;][]{2001ApJ...553..174G, 2003MNRAS.346L..36R}. It has been hypothesized that such self-gravitating over-densities could be precursors to giant planets \citep{1997Sci...276.1836B, 1998ApJ...503..923B}. In fact, formation by gravitational instability is favoured for giant planets on wide orbits \citep[e.g.][]{2019Sci...365.1441M}. This poses the question of whether GoHam~b may be a young protoplanet formed via gravitationally instability.

\begin{figure*}
\includegraphics[]{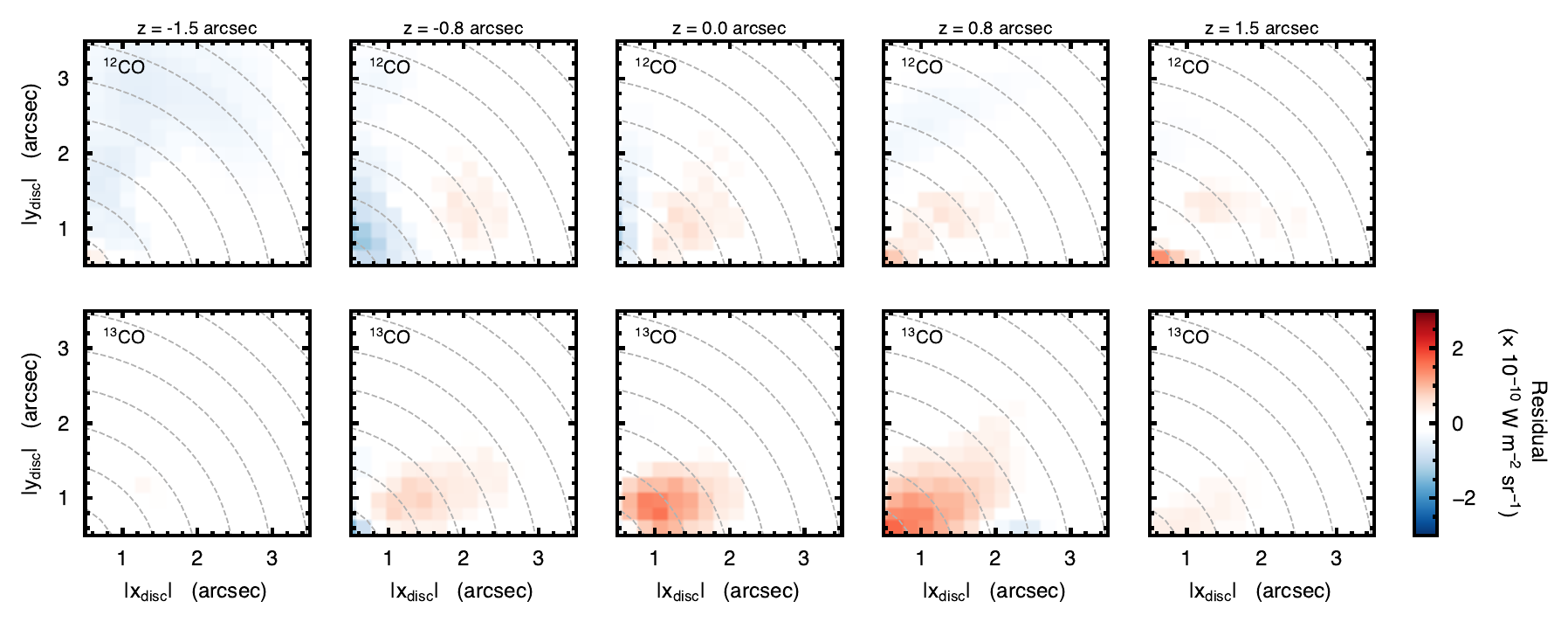}
\caption{Residuals between the positive and negative $x_{\rm disc}$ panels from Figs.~\ref{fig:GoHam_cuts_12CO}, top row, and \ref{fig:GoHam_cuts_13CO}, bottom row. Each column represents a slice at a different height above or below the midplane, given at the top of each column. Positive values represent that the positive $x_{\rm disk}$ side is brighter, while negative values suggest that the negative $x_{\rm disc}$ side is brighter. The location of GoHam~b is more readily seen for the $^{13}$CO emission. \label{fig:GoHam_xyresiduals}}
\end{figure*}

To test this hypothesis, we need to understand the three dimensional structure of the edge-on disc, which can be achieved using the deprojection techniques discussed above. In panels \emph{(c)} through to \emph{(f)} of Figure~\ref{fig:GoHam_cuts_13CO}, the right half of the disc (corresponding to the southern half of the disc on the sky) is considerably brighter than the left half which we interpret as GoHam~b. A similar asymmetry is seen in the $^{12}$CO emission, however at a much lower significance. This is more readily seen in Fig.~\ref{fig:GoHam_xyresiduals} which shows the residuals between positive $x_{\rm disk}$ and negative $x_{\rm disc}$ quadrants of the deprojections shown in Figs.~\ref{fig:GoHam_cuts_12CO} and \ref{fig:GoHam_cuts_13CO}. While this projection leave it ambiguous whether the feature is at positive or negative $x_{\rm disc}$, it is clear from the brighter southern side of the disc that these residuals are dominated by an excess of emission in the positive $x_{\rm disc}$ direction. The deprojection shows that the excess in emission is localized in all three dimensions, further confirming it as a local over-density and not due to chance line of sight projection effects. These properties are consistent with what would be expected from an object formed via gravitational fragmentation of the disc.

GoHam~b is also tentatively detected in the $^{13}$CO panel of Fig.~\ref{fig:GoHam_TRD}, consistent in location with the bright peaks seen in the zeroth moment maps in Figs.~\ref{fig:GoHam_cuts_12CO} and \ref{fig:GoHam_cuts_13CO}. However, without the line-of-sight deprojection discussed above, it is hard to fully disentangle the contribution from GoHam~b relative to the background. Future observations designed for these sort of analyses will benefit from first inferring a disc-averaged $(r,\, z)$ emission distribution, before using that as a background model to more readily identify deviations in the line-of-sight deprojections.

The enhancement in the brightness temperature shown in Fig.~\ref{fig:GoHam_cuts_13CO} is $\approx 20\%$, comparable to that found in previous studies of this source \citep{Bujarrabal_ea_2009}. For an optically thin molecular line, the emission is linearly proportional to the product of the gas temperature and the column density of the emitting molecule, while an optically thick line is only proportional to the gas temperature. It is therefore tempting to assume that $^{13}$CO is optically thin and thus offers a direct probe of the mass of GoHam~b. However, given the unresolved nature of GoHam~b and the lack of multiple transitions to infer the local excitation conditions, see Section~\ref{sec:discussion:discmass}, we do not have sufficient information to improve the estimates made previously regarding the mass of GoHam~b, 0.8 to $11.4~M_{\rm Jup}$. Future observations of multiple transitions of optically thin lines will therefore provide the most accurate probe of the mass of GoHam~b, leading to clues about its nature.

\subsection{Utility in Determining Chemical Stratification}

As previously discussed in \citet{Dutrey_ea_2017}, these deprojection techniques allow us to directly access the vertical stratification of molecular species, an essential data product with which to confront astro-chemical models. This approach is hugely complementary to observations of moderately inclined discs which use the asymmetry of the line emission about the disc major axis in a moderately inclined disc to infer the height of the emission surface \citep[e.g.][]{Rosenfeld_ea_2013, Pinte_ea_2018}.

Firstly, the technique for moderately inclined discs can only be applied to bright lines such that the emission in any given channel is well defined. This criteria leaves only $^{12}$CO and $^{13}$CO as viable choices, meaning that less abundant molecules believed to arise from elevated regions, such as CH$_3$CN \citep{Loomis_ea_2018}, are unable to be tested. Conversely, for an edge-on disc there is no requirement on the significance of the detection; if the molecular emission can be detected in the channel maps, it can be deprojected into disc-centric coordinates.

\begin{figure}
    \centering
    \includegraphics{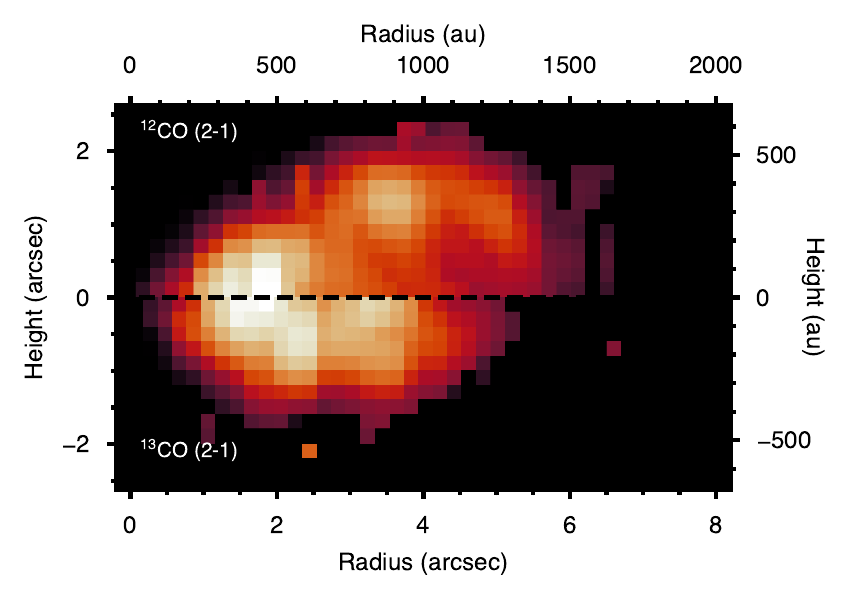}
    \caption{The molecular layers of $^{12}$CO, top, and $^{13}$CO, bottom. Note that the $^{13}$CO has been flipped about $z = 0$ in order to provide a fair comparison due to the asymmetry due to the slight deviation from an edge-on disc. The conversion to linear scales assumed a distance of $d = 250$~pc. Note that the colour scaling has been normalised to the peak value of each to highlight structure.}
    \label{fig:surface_comparison}
\end{figure}

Secondly, the deprojection techniques described in Section~\ref{sec:deprojection} do not require any assumptions about the optical depth of the lines to be made as all pixels can be deprojected to fill in the $(r_{\rm disc},\, z_{\rm disc})$ plane. This can be clearly seen in Figure~\ref{fig:surface_comparison} where the $^{12}$CO emission is detected in the midplane, where usually it is hidden due to high optical depths. For face-on or low inclination discs, an optical depth of 1 is quickly reached along the line of sight such that the disc regions behind this optical surface (the midplane) are hidden from view. Conversely, for an edge-on disc, the line of sight to the disc midplane is unobstructed, allowing us to directly probe the midplane emission without confusion from the upper layers. We additionally note that \citet{Dullemond_ea_2020} showed it is possible access similar information for moderately inclined sources if the spatial resolution of the data allowed for the top and bottom half of the disc to be spatially resolved.

Figure~\ref{fig:surface_comparison} demonstrates these advantages using TRDs of $^{12}$CO and $^{13}$CO emission from GoHam. Note that this figure, unlike Fig.~\ref{fig:GoHam_TRD}, has the colour scales normalised to the peak value for each molecule to bring out the structure of the emission. The $^{12}$CO and likely $^{13}$CO, being optically thick, will be tracing the local gas temperature. In the outer disc, the $^{13}$CO emission may become optically thin, at which point the brightness is proportional to the gas temperature and local CO density. We observe that both molecules peak at elevated regions due to the chemical stratification of the disc rather than an optical depth effect as found in less inclined sources. $^{12}$CO will inhabit a more elevated region than that of $^{13}$CO due to the lower abundance of $^{13}$CO relative to $^{12}$CO resulting in less efficient shielding from photodissociated UV photons. A lower bound for the emission distribution will be given by the freeze-out temperature, $\sim 21$~K for CO. The low midplane temperatures will not completely remove all gaseous-phase molecules, but will significantly reduce their abundance resulting in the very low level of emission seen in Fig.~\ref{fig:surface_comparison}.

These observations demonstrate the utility of edge-on sources in terms of characterising the chemical structures. Moving towards larger samples sizes, observed at higher angular and spectral resolutions will uncover the distribution of molecules currently unable to be constrained with moderately inclined discs. These deprojection techniques are readily combined with line-stacking techniques used to boost the significance of weak lines \citep[e.g.][]{Walsh_ea_2016}, enabling studies of the molecular distribution of weak complex species, studies of which are currently hindered by their lack of bright emission.

\subsection{The Prospect for Mapping the Disc Mass}
\label{sec:discussion:discmass}

With multiple transitions of a molecule observed in an edge-on source, it is possible to go beyond merely mapping out the emission distribution. For example, excitation analyses can be performed to extract local excitation temperatures and volume densities \citep[e.g.][]{Bergner_ea_2018, Loomis_ea_2018, Teague_ea_2018b}. By first deprojecting the data into 3D disc-centric coordinates, one can be certain that the emission being compared arises from the same location; an assumption always made but extremely hard to verify in less inclined sources. In other words, highly-inclined sources provide access to the disc vertical structure, without losing access to the disc azimuthal structure. 

Rarer CO isotopologues are less affected by optical depth issues, and therefore may be more accurate probes of the disc gas mass \citep[e.g. $^{13}$C$^{17}$O][]{Booth_ea_2019}. However, total gas masses derived in this way are sensitive to the assumed CO abundance. With the deprojection, it is also possible to calculate the volume of the emitting area. Thus, if the local $H_2$ density can be constrained using molecules which are not in thermodynamic equilibrium \citep[non-LTE; such as CS in the outer disc, e.g.][]{Teague_ea_2018b}, this can be then be mapped to a total gas mass.

With a gas temperature and local gas mass to hand, it would then be possible to determine whether regions of the GoHam disc are gravitationally unstable \citep[e.g.][]{1964ApJ...139.1217T}. If the region around GoHam b is found to be at (or close to) instability, then this would favour its formation via the gravitational fragmentation of the disc.  Such an interpretation is also supported by recent observations showing that star-disc systems similar to GoHam also appear to be unstable \citep[e.g. HL~Tau][]{Booth_Ilee_2020}.

\section{Summary and Conclusions}
\label{sec:summary}

We have used the deprojection techniques previously presented in \citet{Dutrey_ea_2017}, \citet{Dent_ea_2014}, \citet{Matra_ea_2017} and \citet{Cataldi_ea_2018} to provide a three dimensional view of the massive disc, Gomez's Hamburger using archival SMA observations of $^{12}$CO and $^{13}$CO.

The deprojected data reveals a clear difference between the $^{12}$CO and $^{13}$CO emission regions with the $^{12}$CO tracing a considerably elevated region of $z \, / \, r \sim 0.3$, while the $^{13}$CO arises from much lower regions, $z \, / \, r \lesssim 0.2$, as expected from the higher abundance of $^{12}$CO compared to $^{13}$CO.

When deprojecting the data into the $(x_{\rm disc},\, |y_{\rm disc}|)$ plane, a clear feature in the southern side of the disc in $^{13}$CO which is interpreted as the previously detected over density, GoHam~b. With this deprojection, it is possible to localise the emission to $(r_{\rm disc},\, \phi_{\rm disc}) \approx (500~{\rm au},\, \pm30\degr)$, with the accuracy ultimately limited by the spatial and spectral resolution of the data.

We conclude with a discussion on the utility of these observational techniques in mapping the physical and chemical structure in protoplanetary discs. With access to the full 3D structure of the disc, future observations will be able to map out the gas temperature and density as has never been done before. 

\section*{Acknowledgements}

RT thanks Gianni Cataldi for discussions on unit transformations. We thanks the referee for a helpful and constructive report. RT acknowledges support from the Smithsonian Institution as a Submillimeter Array (SMA) Fellow. TJH is funded by a Royal Society Dorothy Hodgkin Fellowship. JDI acknowledges support from the STFC under ST/R000549/1. MRJ is funded by the President's PhD scholarship of the Imperial College London and the `Dositeja' stipend from the Fund for Young Talents of the Serbian Ministry for Youth and Sport.


\bibliographystyle{mnras}
\bibliography{main}

\bsp	
\label{lastpage}
\end{document}